\documentclass[twocolumn,amsmath,amssymb,aps,prb]{revtex4-1}

\usepackage{times}
\usepackage{amsmath}
\usepackage{amssymb}
\usepackage[]{graphicx}
\usepackage{bm}
\usepackage{color}
\usepackage{soul}
\usepackage{float}
\usepackage{hyperref}
\usepackage{epstopdf}  
\begin{document}

\title{Preferential rotation of chiral dipoles in isotropic turbulence}

\author{Stefan Kramel}
\author{Greg A. Voth}
\email{gvoth@wesleyan.edu}
\homepage{http://gvoth.research.wesleyan.edu/}
\affiliation{Department of Physics, Wesleyan University, Middletown, Connecticut 06459, USA\\
}

\author{Saskia Tympel}
\affiliation{Department of Applied Physics, Eindhoven University of Technology, 5600 MB, Eindhoven, The Netherlands\\}
\author{Federico Toschi}
\email{f.toschi@tue.nl}
\homepage{http://toschi.phys.tue.nl/wordpress/}
\affiliation{Department of Applied Physics and Department of Mathematics and Computer Science, Eindhoven University of Technology, 5600 MB, Eindhoven, The Netherlands\\}
\affiliation{Istituto per le Applicazioni del Calcolo, Consiglio Nazionale delle Ricerche, Via dei Taurini 19, 00185 Rome, Italy\\
}

\begin{abstract}

Particles in the shape of chiral dipoles show a preferential rotation in three dimensional homogeneous isotropic turbulence. A chiral dipole consists of a rod with two helices of opposite handedness, one at each end. We can use 3d printing to fabricate these particles with length in the inertial range and track their rotations in a turbulent flow between oscillating  grids. High aspect ratio chiral dipoles will align with the extensional eigenvectors of the strain rate tensor and the helical ends will respond to the strain field by spinning around its long axis. The mean of the measured spinning rate is non-zero and reflects the average stretching the particles experience. We use Stokesian dynamics simulations of chiral dipoles in pure strain flow to quantify the dependence of spinning on particle shape. 
Based on the known response to pure strain, we build a model that gives the spinning rate of small chiral dipoles using Lagrangian velocity gradients from high resolution direct numerical simulations.  The statistics of chiral dipole spinning determined with this model show surprisingly good agreement with the measured spinning of much larger chiral dipoles in the experiments. 

\end{abstract}

\date{\today}

\maketitle
An incompressible turbulent fluid flow produces exponential stretching of material line segments. In 1952, Batchelor conjectured that this must occur~\cite{1952Batchelor}, and subsequent work has confirmed his conjecture, determining that their average exponential growth rate is $\zeta=\langle e_{i}S_{ij}e_{j}\rangle\approx0.12\tau_\eta^{\scriptscriptstyle{-1}}$, where $S_{ij}$ is the strain rate tensor, $\tau_\eta$ the Kolmogorov time, and $e_i$ is the orientation unit vector~\cite{1990Girimaji,2007Goto,2015Byron}. One might wonder how an incompressible flow can stretch material lines on average since every fluid element must contain contraction to balance extension and maintain constant volume. The answer lies in the Lagrangian advection of material lines which causes them to preferentially orient along extensional directions of the velocity gradient tensor. If material lines are oriented randomly, they have zero mean stretching rate. But after being advected for approximately $10\tau_\eta$, material lines reach a steady state alignment with respect to the velocity gradient tensor, and have a positive mean stretching rate. This problem has recently received attention from experiments and simulations studying small thin rods in turbulence~\cite{2011Pumir,2012Parsa,2015Ni}.  

Valuable insights concerning the development of intermittency in turbulent flows have been obtained by considering the `advected delta-vee' system, in which velocity differences are sampled between two points advected in the flow but constrained to maintain fixed distance between them~\cite{2005Li,2006Li}. If we consider these points are randomly oriented, then the mean longitudinal velocity difference, $\langle\Delta u_r \rangle$, must be zero as a consequence of incompressibility. To obtain insights into the dynamics of turbulence from these longitudinal velocity differences, one needs to consider higher moments. For example, the third moment in the inertial range is related to the mean energy dissipation rate by Kolmogorov's 4/5~law: ${\langle\left(\Delta u_{r}\right)^{3}\rangle = -\frac{4}{5}\langle\epsilon\rangle r}$. However, if the two points are allowed to advect with the flow, they develop a preferential orientation with respect to the velocity gradient tensor.   In this oriented Lagrangian reference frame, the mean velocity difference becomes greater than zero. In particular, for small $r$, the mean velocity difference is ${\langle \Delta u_r \rangle/r=\zeta}$.   

This same process is responsible for vortex stretching in turbulent flows. Vorticity is partly advected by the flow and becomes aligned with the extensional directions of the velocity gradient tensor, leading to a positive mean vortex stretching rate~\cite{1987Ashurst, 1997Tsinober, 1999Ooi}.

\begin{figure}
\begin{center}
\includegraphics[width=3.4in]{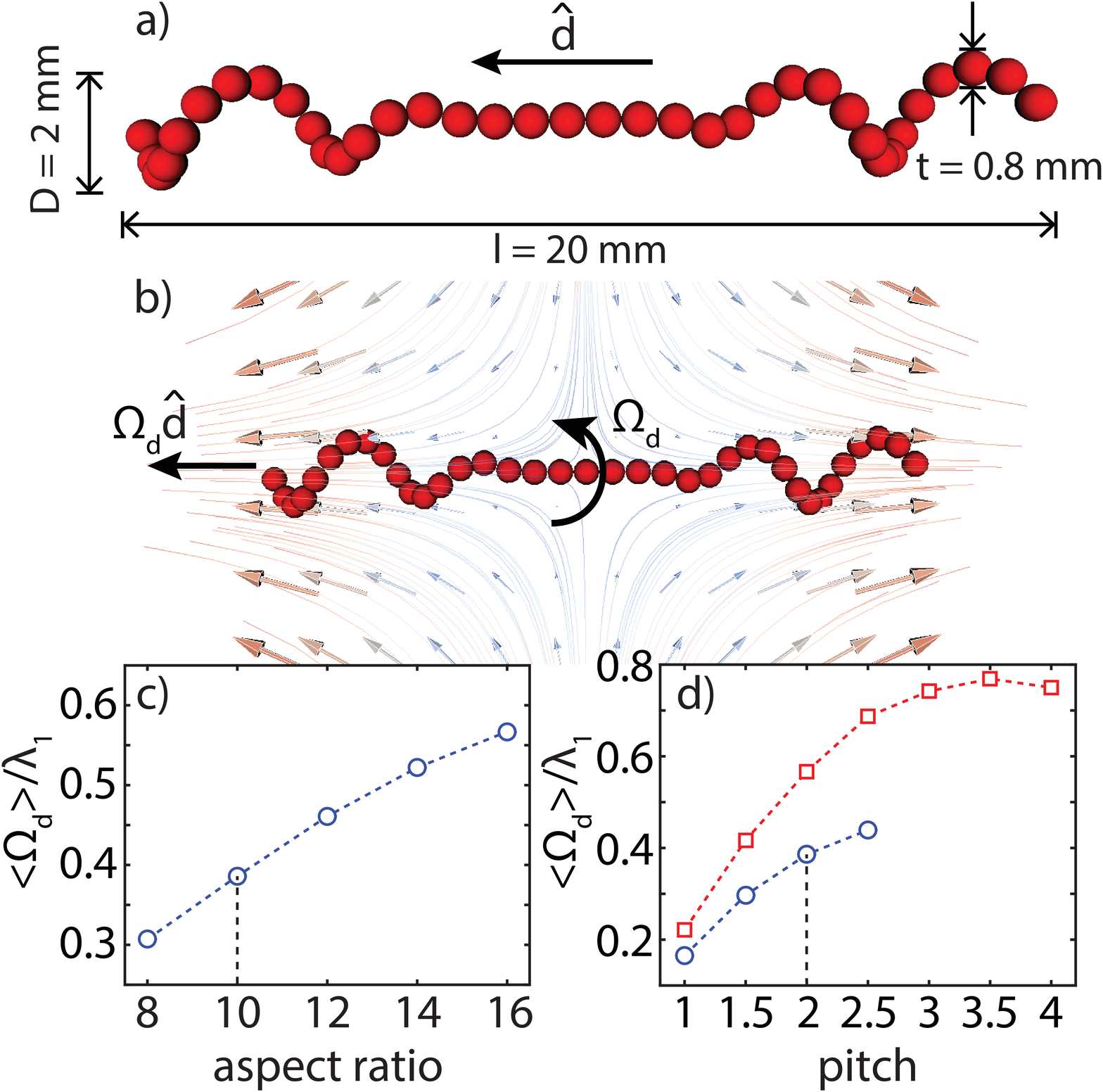}
\caption{\label{fig:aspect_pitch}(a) Model of a chiral dipole. The dipole vector $\hat{d}$ points from right-handed helix to left-handed helix. (b) Response of a chiral dipole to a pure strain flow as indicated by the arrows. (c) Mean spinning rate $\langle \Omega_{d} \rangle$ as function of the overall aspect ratio, $\alpha=l/D$ ($\circ$: pitch $=2$ ). (d) Mean spinning rate as function of helix pitch ($\circ$: $\alpha=10$, $\square$: $\alpha=16$). The mean spinning rate is measured in the Stokesian dynamics simulations and is normalized by the largest eigenvalue of the strain rate tensor $\lambda_{1}$.}
\end{center}
\end{figure}

In this letter, we introduce a new particle design that responds to stretching with a preferential rotation. Measuring rotations of these particles with multiple high speed cameras allows us to experimentally observe the mean stretching experienced by orientable elements in turbulent fluid flow. The particle has two helical ends with opposite handedness as shown in Fig.~\ref{fig:aspect_pitch} (a). We call these particles chiral dipoles because of their similarity to electrical dipoles. The total chirality of the particle is zero; however, the two ends with opposite chirality are separated by a fixed distance. The chiral dipole vector,  $\hat{d}$, points from the right-handed end to the left-handed end. When placed in a pure strain flow, a high aspect ratio chiral dipole like the one shown in Fig.~\ref{fig:aspect_pitch} (b) tumbles until $\hat{d}$ points along the extensional strain direction. The strain flow then couples to the chiral dipole shape to produce a solid body rotation rate in the direction of the chiral dipole vector, $\mathbf{\Omega}=\Omega_d \hat{d}$, where $\Omega_d$ is called the spinning rate. A similar particle design was mentioned by \citet{1997Purcell}, predicting that such a particle should sink without spinning in a quiescent fluid.  

We use Stokesian dynamics simulations \cite{1961Bretherton} to quantify the effects of a particle's shape on its rotational motion and combine the results with direct numerical simulations to study their dynamics in a turbulent flow. It is reasonable to model a complex shaped particle like a chiral dipole with individual spheres (monomers), each fixed in their relative position as shown in Fig.~\ref{fig:aspect_pitch} (a). We used 40 monomers to reproduce the experimental particles while keeping the computational time acceptable. The model is allowed to rotate freely while it is subjected to constant velocity gradients. We can tweak the shape of the model until we get the desired response to the strain flow. 

In the case of chiral dipoles, the most important parameters are the aspect ratio of the particle, $\alpha=l/D$ and the pitch of the helices.  As shown in Fig.~\ref{fig:aspect_pitch} (a), $l$ is the length of the particle and $D$ is the diameter of the helices. The pitch is defined as the length along the helix axis for a complete turn divided by the diameter $D$. We know the particle should have a high aspect ratio, $\alpha\gg1$ to ensure good alignment with the extensional eigenvectors of the strain rate tensor\cite{2012Parsa}.  

Figure~\ref{fig:aspect_pitch} (c) and (d) show the mean spinning rate from Stokesian dynamics simulations of chiral dipoles in a two-dimensional pure strain flow with strain rate eigenvalues $\lambda_1$, $\lambda_2=0$ and $\lambda_3=-\lambda_1$.  After an initial orientation phase of 5 to 10 Kolmogorov times, the particle aligns with the extensional eigenvector of the strain rate tensor and begins to spin about its long axis at a rate $\Omega_d$, with the mean value calculated in this aligned state.   Figure~\ref{fig:aspect_pitch} (c) shows that increasing the aspect ratio with constant pitch increases the spinning of a chiral dipole in a strain flow. Figure~\ref{fig:aspect_pitch} (d) shows the mean spinning rate as a function of pitch with constant aspect ratio and suggests that there is an optimal pitch.  Theoretically, a particle with pitch near 3.5 and very high aspect ratio would yield the largest coupling of spinning to the strain rate in experiments.  However, our 3D printer did not yield structurally stable particles with smallest dimension less than $t=0.8$ mm, and we need $D\gg t$ in order to allow optical reconstruction of the particle's 3d orientation. An interesting question for future research would be to determine the optimal shape for coupling spinning to the fluid flow in both pure strain and in turbulent flows.

\begin{figure}
\begin{center}
\includegraphics[width=3.4in]{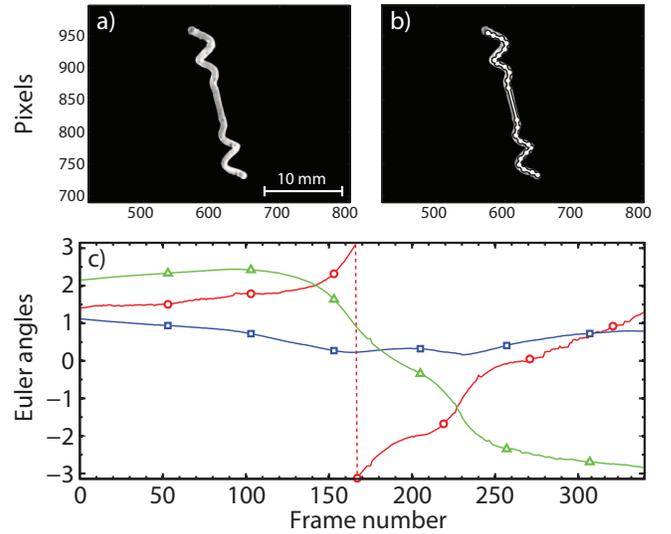}
\caption{\label{fig:model-fit} (a) Cropped image of a chiral dipole from one camera. (b) Projection of the model onto the image plane of the camera using the measured Euler angles. (c) Time series of experimentally measured Euler angles ($\bigtriangleup=\phi$, $\square=\theta$, $\circ=\psi$) of a chiral dipole along its trajectory.} 
\end{center}
\end{figure}

The experiments were performed in a turbulent flow between oscillating grids~\cite{2010Blum}. The grids were driven in phase at a frequency of 1 Hz and 3 Hz in separate runs, resulting in a Taylor Reynolds number of $R_{\lambda} = 120 $ and $R_{\lambda} = 183 $, respectively.  The parameters characterizing the turbulent flow are shown in Table~\ref{tab:exp_param}. We use 3D printing technology~\cite{2014Marcus} to fabricate 2000 chiral dipoles with aspect ratio $\alpha = 10$, pitch 2 and a largest dimension of 20 mm, which corresponds to 35$\eta$ and 72$\eta$, depending on the Reynolds number.  Spherical tracer particles with a diameter of 150 $\mu$m were used to measure the rms fluid velocity and to calculate the third order longitudinal structure functions from which we determine the energy dissipation rate.   In order for the chiral dipoles to be neutrally buoyant, the fluid was density matched by adding CaCl$_2$ until a fluid density of 1.20 g cm$^{-3}$ was reached. This resulted in a fluid viscosity of $\nu = 2.00$ mm$^2$ s$^{-1}$. The particles are fluorescent and illuminated with laser beams from four directions to minimize self-shadowing \cite{2014Marcus}. Four cameras image the particles from different angles at a frame rate of 450 Hz. Using the images and camera calibration parameters from all four cameras, we can measure the three Euler angles defining the orientation of the chiral dipole. This is done by projecting a model of the particle onto the image plane of each camera and letting a non-linear least squares search find the Euler angles that minimize the difference between the projected model and the data. The model consists of 30 connected rods (the endpoints of each rod are represented by circles in Fig.~\ref{fig:model-fit} (b). Figure ~\ref{fig:model-fit} (c) shows the measured Euler angles along a typical trajectory.  The full solid-body rotation vector can be measured by fitting the particles orientation over several time steps along individual trajectories. The model can easily be adjusted to match other particle shapes and the algorithm can extract the orientation.

In addition to the experimental measurements, we use direct numerical simulations (DNS) of homogenous isotropic turbulence at a Reynolds number of $R_{\lambda} = 400 $ to calculate the motion of a chiral dipole along its trajectory. The simulation volume includes a total of ${N^3 = 2048^3}$ collocation points and $\mathcal{O}(10^7)$ measurements of Lagrangian velocity gradients for a few large eddy turnover times \cite{2009Benzi}. The characteristic quantities of the simulations are summarized in Table~\ref{tab:exp_param}.  

\begin{table}
\caption{\label{tab:exp_param}Flow parameters: $R_{\lambda}=(15\bar{u}L/\nu)^{1/2}$ Taylor Reynolds number, $L=\bar{u}^{3}/\epsilon$ energy input length scale, $\bar{u}=(\langle u_{i}u_{i}\rangle/3)^{1/2}$ rms-velocity, $\epsilon$ mean energy dissipation rate, $\eta=(\nu^{3}/\epsilon)^{1/4}$ Kolmogorov length scale, $\tau_{\eta}=(\nu/\epsilon)^{1/2}$ Kolmogorov time scale, $\nu=2.00 \times 10^{-6}$ m$^2$ s$^{-1}$ kinematic viscosity.}
\begin{ruledtabular}
  \begin{tabular}{ccccccc}
		\multicolumn{7}{c}{Experiments} \\
    Grid freq. & $R_{\lambda}$ & $L$ & $\bar{u}$ & $\epsilon$  & $\eta$& $\tau_{\eta}$ \\
		\lbrack Hz\rbrack &  & [mm] & [mm s$^{-1}$] &  [mm$^2$ s$^{-3}$] & [mm] & [s] \\
		\hline 
       1   & 120 & 94 & 20.4 & 90 & 0.546 & 0.149 \\
       3   & 183 & 80 & 55.6 & 2150 & 0.247 & 0.030 \\
			\\
		\multicolumn{7}{c}{Simulations} \\
		$N$ &  $R_{\lambda}$ & $L$ &  $\bar{u}$ & $\epsilon$ & $\eta$ & $\tau_{\eta}$ \\
		\hline
		$2048$ &  400 & 4.08 & 1.411 & 0.687 & 0.0028 &  0.0225 \\
  \end{tabular}
\end{ruledtabular}
\end{table}

High aspect ratio chiral dipoles can be approximated by rods and their tumbling rate can therefore be described by Jeffery's equation \cite{1922Jeffery}

\begin{equation}
\dot{d}_{i} = \Omega_{ij}d_{j} + \frac{\alpha^2\!-\!1}{\alpha^2\!+\!1}\!\left(S_{ij}d_{j}\!-\!d_{i}d_{k}S_{kl}d_{l} \right) 
\end{equation}

where $S_{ij}$ (strain rate tensor) and $\Omega_{ij}$ (rotation rate tensor) are the symmetric and anti-symmetric parts of the velocity gradient tensor, respectively.  We can use the velocity gradients from the DNS to integrate Jeffery's equation and obtain the orientation of a particle that has been aligned by the flow. In addition to tumbling, a thin rod is also spinning around its symmetry axis with half of the fluid vorticity $\boldsymbol{\omega}$ in that direction.  It is reasonable to assume the spinning rate of chiral dipoles along $\hat{d}$ has an additional contribution which comes from the strain flow

\begin{align}
\Omega_{d} &= \frac{\omega_{i}}{2} d_{i} + \beta~d_iS_{ij}d_j \\
					 &= A_{S}  + \beta A_{L} 
\label{eq:spinning}
\end{align}

The constant $\beta$ is dependent on the particle shape and describes the strength of the coupling of the spinning rate to the strain field. An approximate value for $\beta$ for our particle shape was obtained from the Stokesian dynamics simulations, where $\beta = 0.39$. We adopt the compact notation developed for the analysis of the `advected delta-vee' system by \citet{2005Li,2006Li} to define the longitudinal and transverse velocity gradients with respect to the particle. The longitudinal component is $A_L\!=\!d_iS_{ij}d_j$ and the transverse component is the magnitude of the tumbling rate $A_N\!=\!(\dot{d}_{i}\dot{d}_{i})^{1/2}$. We can complete the picture if we include the spinning due to the fluid vorticity $A_{S}\!=\!\frac{1}{2}\omega_{i}d_{i}$. These quantities provide a good description of the rotation of a chiral dipole. One can immediately see that chiral dipoles can be used to measure $A_L$ experimentally, if the fluid vorticity is known.

\begin{figure}
\begin{center}
\includegraphics[width=3.4in]{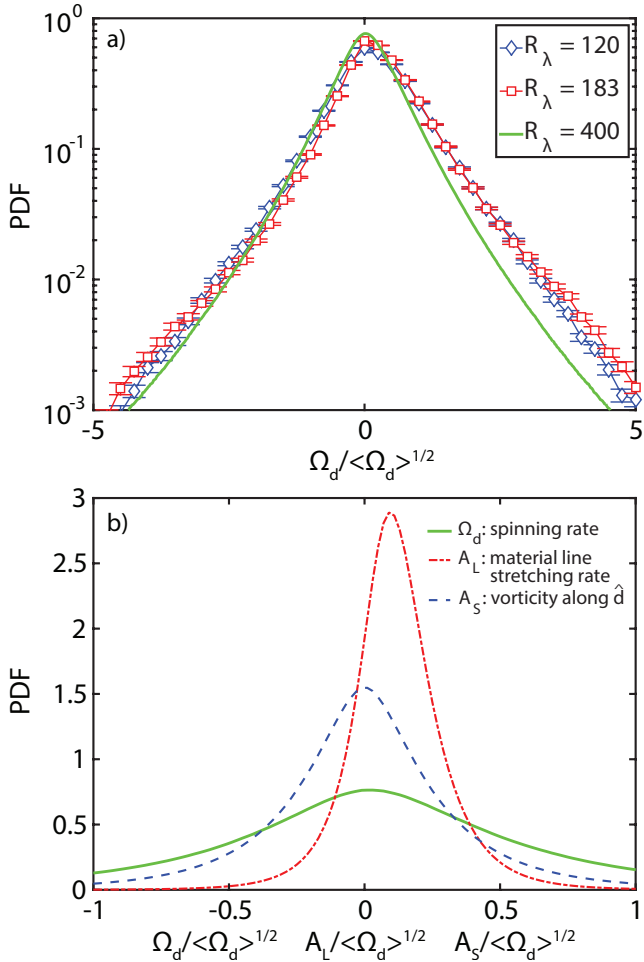}
\caption{\label{fig:pdf_spinning}(a) Probability density function (PDF) of the spinning rate $\Omega_{d}$ normalized by the standard deviation for both Reynolds numbers (blue $\diamond$, ${R_{\lambda}\!=\!120}$ and red $\circ$, ${R_{\lambda}\!=\!183}$) and simulations (green solid line). (b) PDF of the individual contributions from strain, $A_L\!=\!d_iS_{ij}d_j$ (red dashed-dotted line) and vorticity, $A_S\!=\!\frac{1}{2}\boldsymbol{\omega}\!\cdot\!\hat{d}$ (blue dashed line) to the spinning rate $\Omega_{d}$ (solid green line). The standard deviation of the simulations is $\langle\Omega_{d}^{2}\rangle^{1/2}=0.85 \tau_{\eta}^{-1}$.}
\end{center}
\end{figure}

Figure~\ref{fig:pdf_spinning} (a) shows the probability density function (PDF) of the spinning rate from both experimental measurements and the simulations.  There is a clear asymmetry around zero, with a larger probability of positive spinning rate.  This is clear evidence of the preferential rotation direction of chiral dipoles advected in isotropic turbulence.   In the simulations, we can separate the contributions from strain and vorticity as shown in Fig.~\ref{fig:pdf_spinning} (b).  Since a chiral dipole is equally likely to be parallel or anti-parallel to the vorticity vector, its mean contribution is zero. The contribution from the strain is responsible for the non-zero mean spinning rate. 
 
The shape of the experimentally measured PDFs in Fig.~\ref{fig:pdf_spinning} (a) depends fairly strongly on the fit length used to measure the solid body rotation rate. Shorter fit lengths include more noise from the orientation measurements, leading to larger tails, whereas longer fit lengths filter out events of large rotational acceleration. Both experimental curves in Fig.~\ref{fig:pdf_spinning} have been measured with a fit length of $0.5\tau_{\eta}$. The PDFs collapse surprisingly well given the fact that the experiments were performed with particles in the inertial range and the simulations are for particles in the dissipation range.

The experimentally measured mean spinning rate is ${\langle\Omega_d\rangle = 0.58\tau_{l}^{-1}}$ (${\langle\Omega_d\rangle = 1.08\tau_{l}^{-1}}$) for ${R_\lambda=120}$ (${R_\lambda=183}$) normalized by ${\tau_l = l/u_l}$, where ${u_l=\langle\left(\Delta u_l\right)^2\rangle^{1/2}}$ is the magnitude of the longitudinal velocity difference at separation $l$.  We see that a simple scaling law with the mean spinning rate scaling like the coarse grained velocity gradient does not hold.  The larger than expected spinning rate of the larger chiral dipoles may be explained by two factors.   First, the preferential alignment between the particle orientation and the extensional eigenvectors of the coarse grained strain rate tensor likely depends on particle size. \citet{2007Luthi} measured the coarse grained velocity gradient tensor and showed that the preferential alignment of vorticity moves toward the maximum extensional eigenvector as the coarse graining length scale increases. Second, the coupling constant $\beta$ may depend on the particle Reynolds number. It is possible that chiral dipoles spin more efficiently in a turbulent environment than in the Stokes flow limit. Future work using numerical simulations of particles with lengths in the inertial range and experiments using particles with lengths at the Kolmogorov scale could clarify how the crossover from dissipation to inertial range scales affects the rotations of chiral dipoles.  

\begin{figure}
\begin{center}
\includegraphics[width=3.0in]{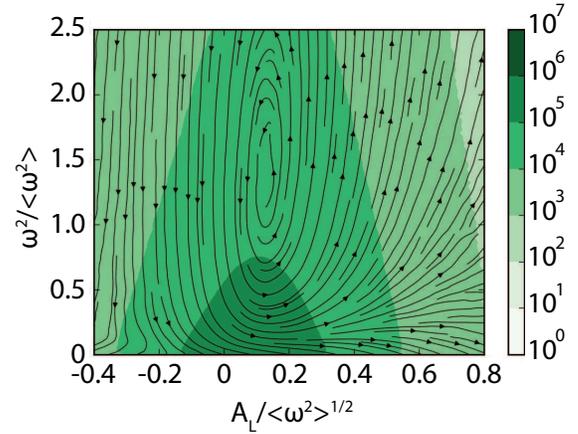}
\caption{Mean trajectories are representing the cyclic behavior of fluid elements in the phase space spanned by enstrophy, $\boldsymbol{\omega}^2$ and material line stretching rate $A_L$ from the DNS. The color map shows the PDF.}
\label{fig:QW}
\end{center}
\end{figure}

Stretching of material lines is one of the fundamental processes in the energy cascade.  Traditionally, vortex stretching has been emphasized with more recent work highlighting that strain-strain interactions are equally if not more important.  The ability to follow elongated particles through the flow and observe the preferential stretching they experience suggests new ways to quantify the dynamic processes of the cascade.  Figure~\ref{fig:QW} shows the mean trajectories of fluid elements in the space of enstrophy, $\omega^2$, and the material line stretching rate, $A_L$.  There is a clear cyclical pattern with a fixed point at large enstrophy and a positive value of the material line stretching rate.  A qualitatively similar cycle has been observed for the vortex stretching process by \citet{1999Ooi} reflecting the similar physics involved in vortex stretching and material line stretching.   

Chiral dipoles experience a preferential rotation direction in isotropic turbulence. The ability to fabricate particles with complex shapes and measure their rotational motion opens the door to the study of a wide variety of particle shapes beyond the axisymmetric ellipsoids that have been the focus of most previous work. The mechanism of the preferential rotation is alignment of the slender particles by the fluid strain at the scale of the particle so that the particles experience extensional strain on average which produces preferential rotation due to the chiral ends. These measurements highlight the importance of analyzing turbulent flows in an oriented Lagrangian reference frame~\cite{2005Li,2006Li}. Future work is needed to clarify the scale dependence of preferential alignment and rotation. Study of coarse grained rotation and deformation have yielded substantial insights into the dynamics of turbulence~\cite{2011Xu, 2007Luthi}. Our current experiments and simulations show agreement in the shape of the spinning rate PDF, but the experiments are limited to inertial range particle sizes and the simulations are limited to dissipation range particle sizes. Tools to measure and simulate particles across the full range of turbulent scales could provide a powerful new way to analyze the dynamics of the turbulent cascade process.

We acknowledge support from  NSF grants DMR-1208990 and DMR-1508575. We thank Brendan Cole for assistance with data acquisition, and Guy Geyer Marcus, Rui Ni, Tom Powers, and Luca Biferale for stimulating discussions.

\end{document}